\documentclass[12pt]{article}
 
\input{epsf.tex}
\begin{document}
\begin{center}
{\bf PARAMETERS OF CASCADE GAMMA-DECAY OF COMPOUND-NUCLEI $^{146}$Nd, $^{156}$Gd, $^{172}$Yb, $^{182}$Ta, $^{184}$W, $^{191}$Os,
$^{231,233}$Th, $^{239}$U, $^{240}$Pu 
FROM EXPERIMENTAL DATA OF REACTION $(\overline{n},\gamma)$}\\\end{center}
\begin{center}
 {\bf A.M. Sukhovoj, V.A. Khitrov}\\\end{center}
\begin{center}
{\sl Joint Institute for Nuclear Research, 141980, Dubna, Russia }\\
\end{center}
\begin{abstract}
Re-analysis of experimental data on primary gamma-transitions averaged
over some energy intervals of neutron resonances has been performed.
Approximation of their cumulative sums together with extrapolation of the
obtained distribution to zero value allowed us to determine mean intensities
of of E1- and M1-transitions, their probable number and total dispersion
of intensity deviations from the mean value. The level density and sum
of radiative strength functions determined in this way confirm main
peculiarities  of these nuclear parameters determined from intensities
of the two-step gamma-cascades.
\end{abstract}

\section{Introduction}\hspace*{16pt}
Level density $\rho$ and radiative strength functions
$k=\Gamma/(E_\gamma^3 D_\lambda A^{2/3})$ of the dipole primary transitions
of the neutron resonance gamma-decay provided us with a considerable portion
of experimental information on both nuclear properties on the whole and
nuclear resonances -- in particular.

Nevertheless, there is an urgent necessity in determination of these data in
new independent experiments. The ground for this is principle incompatibility
of experimental data on $\rho$ and $k(E1)+k(M1)$ obtained in two-step reaction
$(n,2\gamma)$ \cite{Meth1,PEPAN-2005} with analogous data from one-step
reactions like $(p,n)$ \cite{Vona83,Zh}, $(d,p\gamma)$ \cite{AdNP}
and $(^3$He,$\alpha\gamma)$ \cite{NIM}.

Partial analysis of possible reasons for this discrepancy was performed,
for example, in \cite{main}. It results in the following statements: 

a) transfer coefficients of total errors in determination of partial
cross-sections of two-step reactions onto errors of the parameters under
determination are much less than those determined in one-step reactions
like $(d,p\gamma)$, $(^3$He,$\alpha\gamma)$ due to other form of functional
dependence between subjects ``spectrum" and ``parameters";

b) the same concerns sensitivity of the determined parameters to degree
of erroneous  of hypothesis by Bohr-Mottelson \cite{BM}
(Axel-Brink \cite{Axel,Brink})for gamma-quanta) of independence of interaction
cross-section of final reaction product in reverse reaction with the
excited final nucleus. 

In practice, error in absolute normalization of the two-step spectra
by order of $\pm 25\%$ changes the $\rho$ and $k(E1)+k(M1)$
values not more than by a factor of two \cite{TSC-err} near
$E_{ex}\approx 0.5B_n$.
This is the largest systematical error in these experimental data.
Analogous error of the total gamma-spectrum normalization for
different excitation energies
$0 \geq \Delta S/S \leq 1\%$ in one-step
(according to the used analysis method) reaction
$(d,p\gamma)$ or $(^3$He,$\alpha\gamma)$ brings \cite{Osloerr} to more than
100\% errors in intensities of all the spectra of only primary gamma-transitions,
at least, for $E_\gamma \leq 0.5$ MeV. 
The coefficients of further transfer of the indicated error on the
determined values of the parameters, until now, were not determined by anyone. 
It is possible to suggest in this situation, that
$\rho$ and $k(E1)+k(M1)$ determined according to \cite{NIM} have arbitrary
systematical errors for different excitation and primary
gamma-transitions energies.

Direct experimental verification of hypotheses \cite{BM,Axel,Brink} is
impossible. But, the possibility of obtaining functionals directly depending
on unknown partial cross-sections of reaction for excited target-nucleus was
found in \cite{PEPAN-2005}. Cascade population of levels determined there for
the majority of $\approx 20$ nuclei up to excitation energy of
$E_{ex} \sim 3-5$ MeV
cannot be reproduced in frameworks of Axel-Brink hypotesis.
However, they can be easily enough reproduced under assumption of existing
enhancement of primary and secondary gamma-transitions to the region of
``step-like" structure in level density \cite{Meth1,PEPAN-2005}.

Unfortunately, complete notion of the observed by this method function
$k(E1)+k(M1)=f(E_\gamma,E_{ex})$ cannot be obtained due to lack of
experimental data. But, as it is seen from comparison between the results
\cite{Yb174,Gd159} and the data \cite{Meth1,PEPAN-2005}, violation of
hypothesis \cite{Axel,Brink} is considerably larger than it was obtained in
\cite{PEPAN-2005}.
For this reason, level density obtained in \cite{Meth1,PEPAN-2005} can be
overestimated by several times in excitation region $\sim 0.5B_n$.
Most probably, this error gradually decreases at lower and higher excitation
energy of nucleus under condition that the experimental data on density
of low-lying levels and neutron resonances have significantly lesser errors.
Strength functions $k(E1)+k(M1)$ are, most probably, underestimated.

Due to this reason, authors \cite{Meth1,PEPAN-2005} performed independent
model-free re-analysis of the experimental data from the $(\overline{n},\gamma)$
reaction in frameworks of only mathematical statistics with the least number
of assumptions on parameters of small sets of the primary gamma-transition
partial widths. It was obtained that the refusal from main postulates of
``statistical" theory brings to conclusion which confirms main results of
\cite{Meth1,PEPAN-2005}.

\section{Main principles of analysis}\hspace*{16pt}

Capture of neutrons in ``filtered" beams, for example, noticeably enough
averages the gamma-transition partial widths in local resonances.
This is true for nuclei with small enough spacing $D_\lambda$ between neutron
resonances. It is possible to observe experimentally the width $\Gamma$ with
relative statistical error $\sigma$ if its value exceeds practically constant
detection threshold $L$ of experiment (in any given narrow interval of
gamma-transition energy).

The portion of primary gamma-transitions of the same multipolarity  and
practically equal energy with $\Gamma <L$ is determined by concrete form
of deviation distribution of $\Gamma$ from mean value in individual
resonances and their effective number for neutron beam in  experiment.

It follows from main statements of modern nuclear theory that the amplitude
of gamma-transition between neutron resonance and low-lying level is determined
by quasiparticle and phonon components in wave functions of both levels
(see, for instance, \cite{Malov}). Their concrete values are determined by
fragmentation degree of the states like $n$ quasiparticles and $m$ phonons
over nuclear levels at different excitation energy. In accordance with
\cite{MalSol}, this process is rather specific -- strength of the fragmented
state is distributed very irregularly. In many cases, its strength is
fragmented over the levels lying near the initial
position of non fragmented state.

In practice, this means that the $\Gamma$ values must strongly and locally
depend on structures of decaying and excited levels.
Their dispersion relative to the average must be determined by number and
value of the wave function components of these levels.
Therefore, fluctuations  of $\Gamma/<\Gamma>$ cannot be described by universal
distribution. Its deviation from the generally adopted Porter-Thomas
distribution \cite{PT} must be determined in every case experimentally.
Practically, it is adopted in analysis (see \cite{Yb174}) that the sum
of dispersions of experimental statistics uncertainty and
``nuclear fluctuations" is equal to $\sigma^2=2/\nu$ with unknown parameter
$\nu$.

The second assumption of analysis is that the gamma-transitions of the same
multipolarity with energy of about some hundreds keV have the same mean value.
In principle, this assumption can be mistaken and, for instance, mean
widths of all the gamma-transitions involved in the set under analysis can
belong to some rather wide interval of possible values. Moreover, the probability
of given mean value may increases as decreasing $<\Gamma>$.
 
This possibility is to be investigated experimentally. Real width distribution
around the average determines reliability of both data presented below and
conclusions of \cite{Meth1,PEPAN-2005}.

\section{The most reliable values}\hspace*{16pt}
The values of level density and radiative strength functions of primary
gamma-transitions obtained by analogy with \cite{Yb174} are presented
in figs. 1--4. The distribution of cumulative sums of reduced intensities for
the analyzed nuclei \cite{Nd146} - \cite{Pu240} and parameters of approximating
curve have no principle
difference with that given in \cite{Yb174,Gd159,U237}.
Therefore, they are not presented in this work.
\begin{figure} %[tbp]
\begin{center}
\leavevmode
\vspace{4cm}
\epsfxsize=17cm
\epsfbox{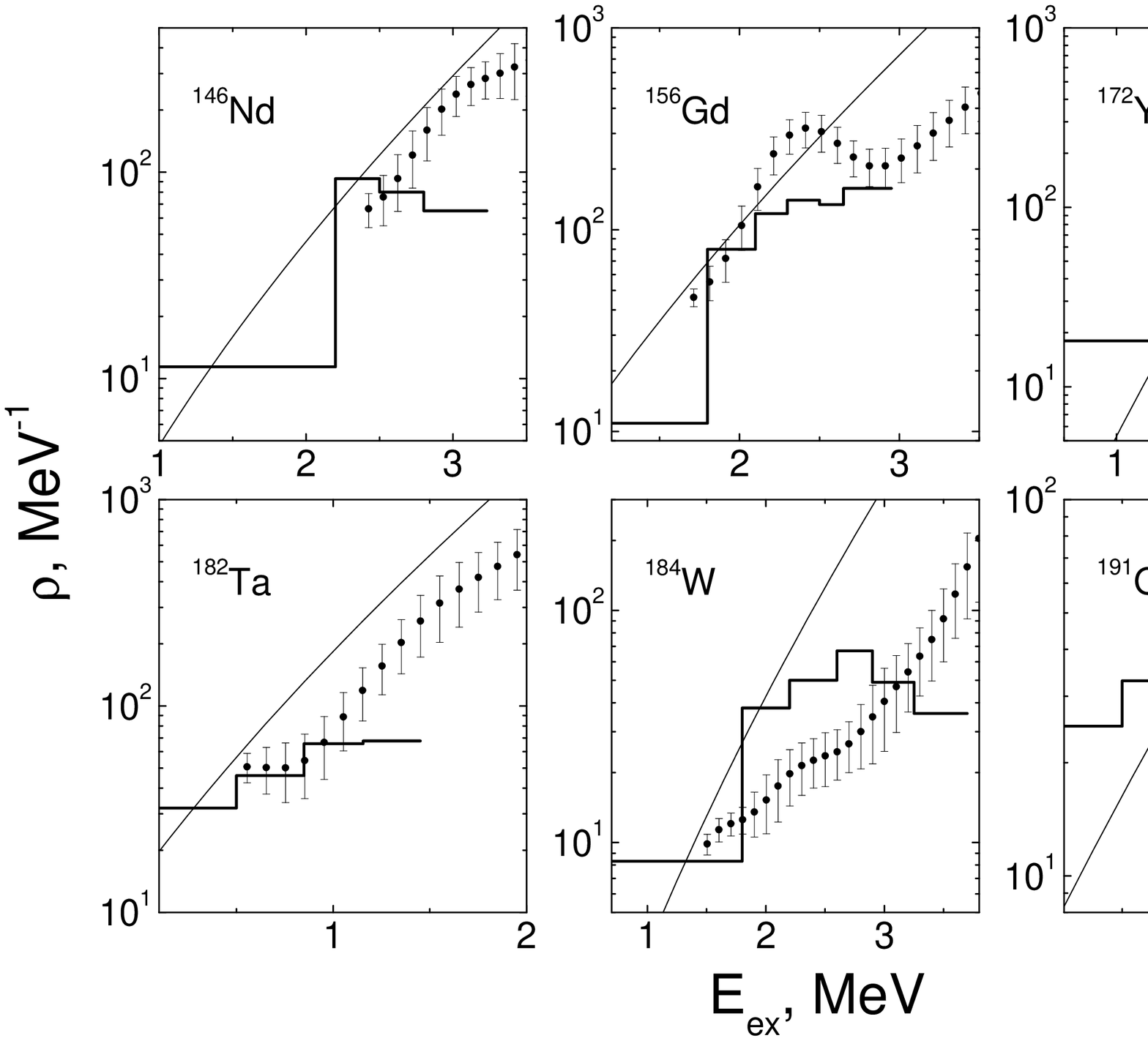}
\end{center}

\vspace{-5.5cm}

{\bf Fig. 1.}  Comparison of different data on level density for the 
$^{146}$Nd, $^{156}$Gd, $^{172}$Yb, $^{182}$Ta, $^{184}$W and $^{191}$Os nuclei.
Curve represents the calculated within model \cite{Dilg}  density of levels
populated by the primary gamma-transitions. Results of presented analysis
-- histogram, points with errors -- data \cite{Meth1}  and \cite{PEPAN-2005}.
\end{figure}

%\newpage

\begin{figure} [tbp]
\vspace{2.5cm}
\begin{center}
\leavevmode
\epsfxsize=12cm
\epsfbox{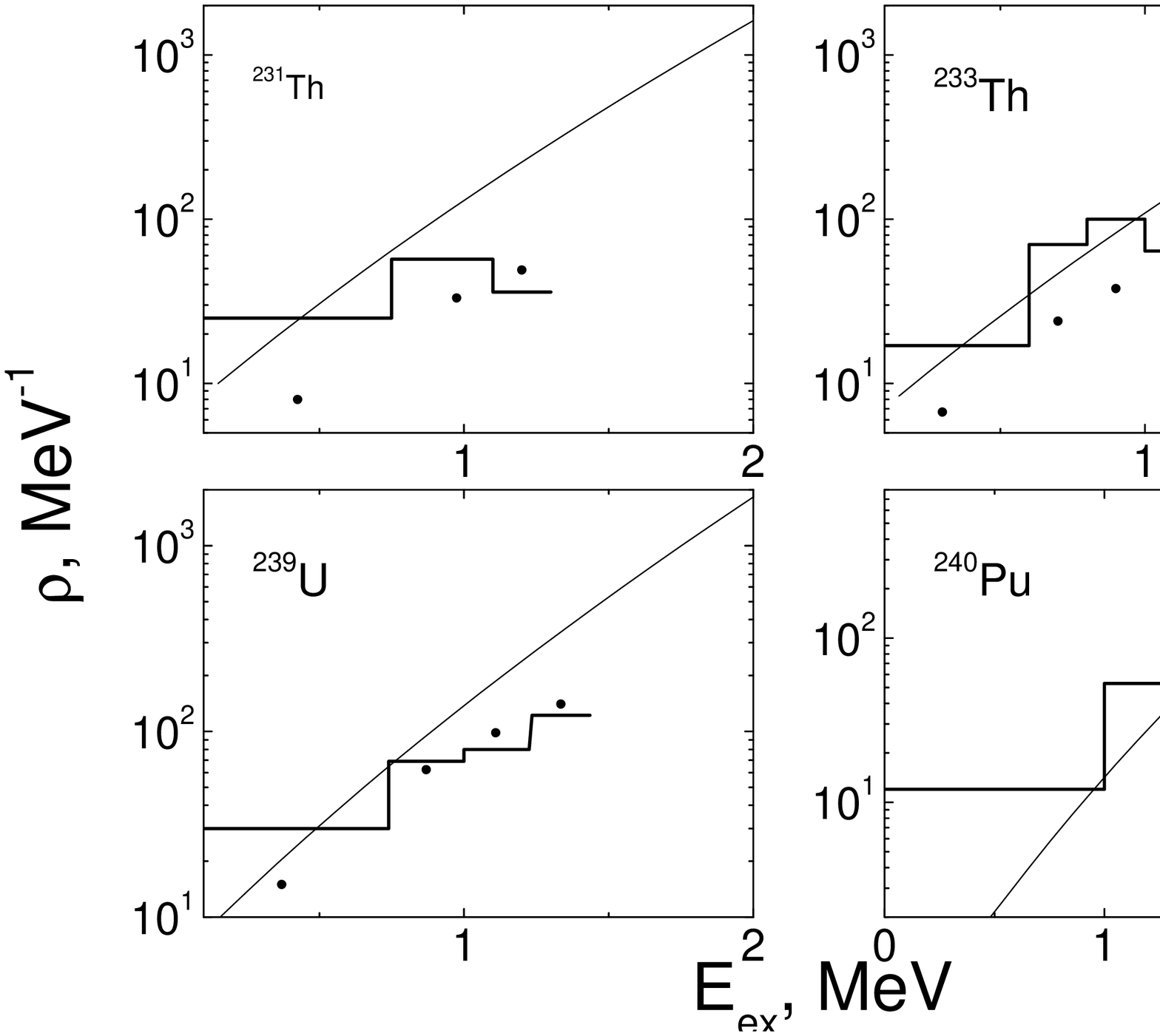}
\end{center}
\vspace{-3.5cm}

{\bf Fig.~2.} The same, as in Fig. 1, for $^{231,233}$Th, $^{239}$U and
$^{240}$Pu. Points show approximation of the experimental data by density
of two- or three-quasiparticle levels of model \cite{Strut} with the
independent on excitation energy coefficient of collective enhancement. 
\end{figure}
\begin{figure} [tbp]
\vspace{3cm}
\begin{center}
\leavevmode
\epsfxsize=17cm
\epsfbox{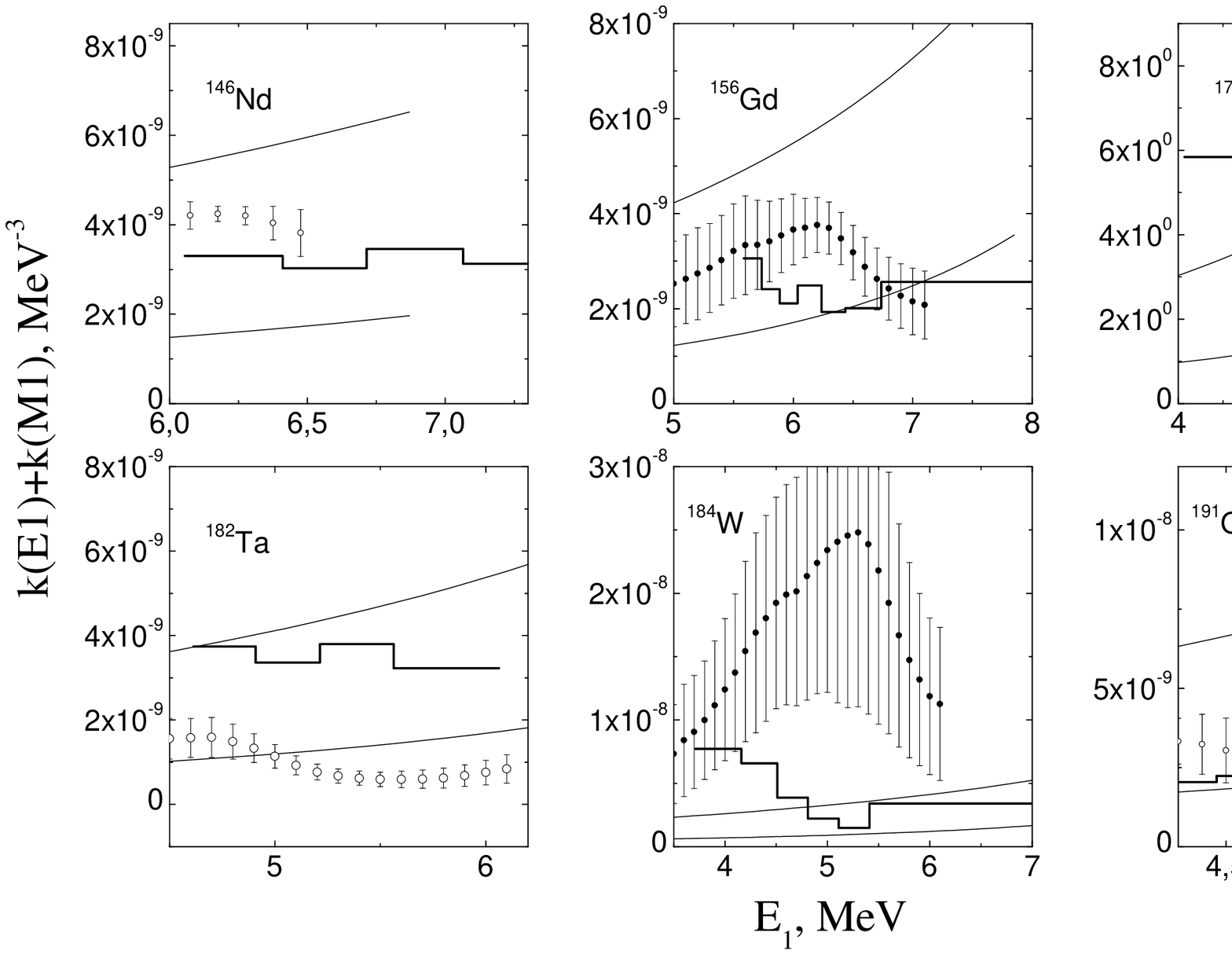}
\end{center}
\vspace{-5.7cm}

{\bf Fig.~3.} The same, as in Fig. 1, for sums of the radiative strength
functions. Results of analysis performed in this work are presented as
histogram of relative values. Upper curve - \cite{Axel}, lower curve -
\cite{KMF} together with $k(M1)$=const.
Points with errors show the data \cite{Meth1,PEPAN-2005}.
\end{figure}

\begin{figure} [htbp]
\vspace{1 cm}
\begin{center}
\leavevmode
\epsfxsize=12cm
\epsfbox{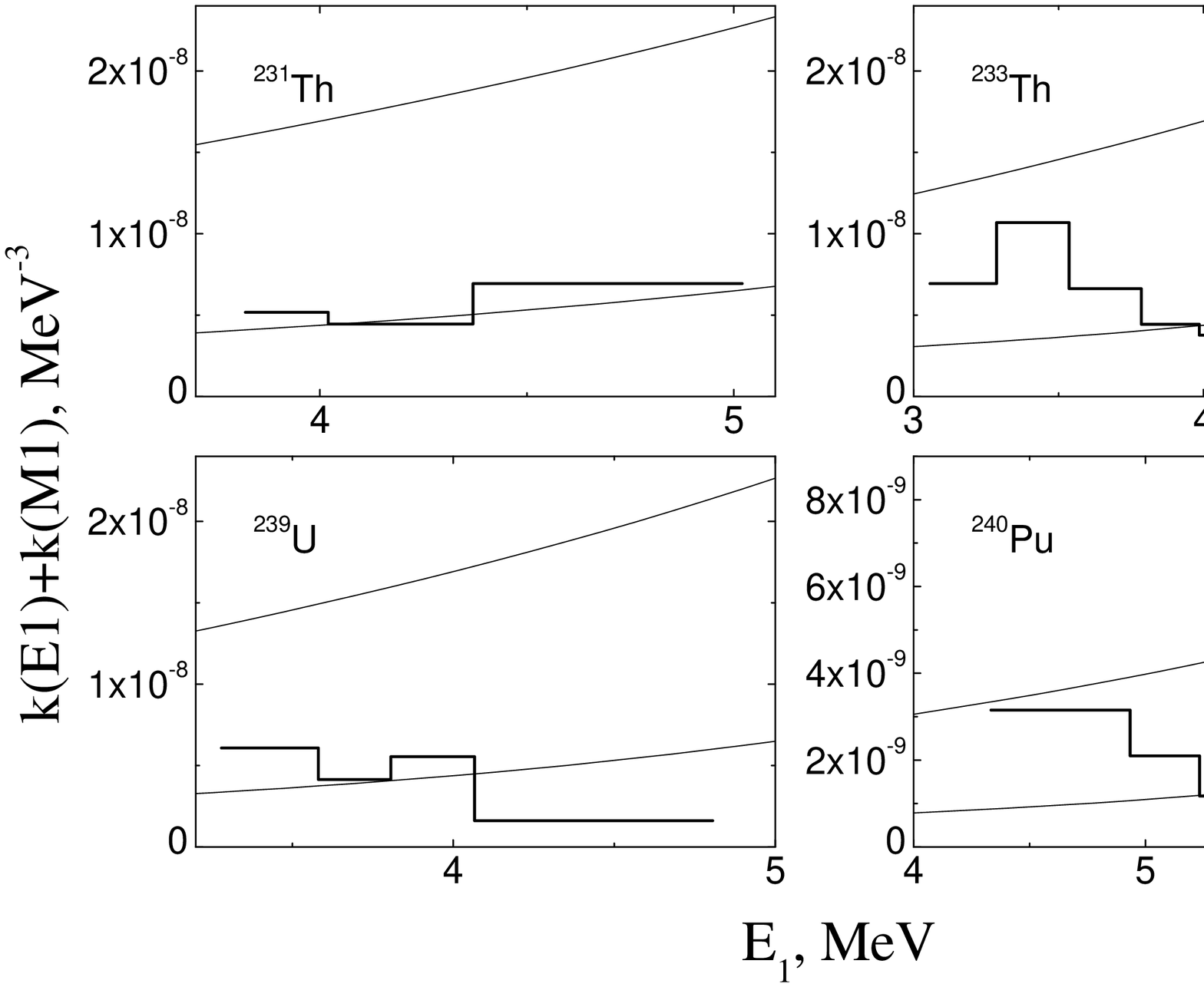}
\end{center}
\vspace{-3.7cm}

{\bf Fig.~4.}  The same, as in Fig. 3, for $^{231}$Th, $^{233}$Th, $^{239}$U
and $^{240}$Pu.
\end{figure}

The gamma-decay parameters of nuclei determined in \cite{Meth1,PEPAN-2005} are
compared with results of two different in principle methods of analysis.
Accounting for results of theoretical analysis \cite{MalSol}
(possible dependence of the wave function components of levels determining
$\Gamma$ on energy of neutron resonances) one can conclude that the discrepancy
between results of two methods of analysis mentioned above is less than their
difference from the data \cite{Zh,NIM}.

It follows, first of all, from observation of ``step-like" structure in level
density in both methods \cite{Meth1,PEPAN-2005,Yb174} and close
to zero or negative derivative
$d(k(E1)+k(M1))/dE_1$ for the same excitation region of nucleus.
Id est, the local peak in sum of strength functions must be more or less
clearly expressed.

The data presented in figs. 1, 2 confirm also conclusion on probable local
increase in density of vibration type levels in the region of the nucleon
pairing energy for a nucleus of a given mass.
The shape of this dependence is presented in details in
\cite{Yb174,Gd159,U237}.

\section{Conclusion}\hspace*{16pt}

Unfortunately, energy interval of the primary gamma-transitions observed in
experiment is considerably less than that for $^{157,159}$Gd, $^{174}$Yb and
$^{237}$U. Nevertheless, the data shown in figs. 1-4 bring to the following
conclusion: notions of a nucleus as a system of non-interacting Fermi-gas and
ideas of other analogous nuclear models \cite{RIPL} are insufficient for
reproduction of modern experimental data.

\end{document}